\documentclass[12pt]{iopart}

\def\binom#1#2{\mathrm{C}({#1},{#2})}

\usepackage{iopams,graphicx}
\begin{document}

\title[Calculation of 1RSB transition temperature under the replica symmetric ansatz]{Calculation of 1RSB transition temperature of spin glass models on regular random graphs under the replica symmetric ansatz}

\author{Masahiko Ueda and Shin-ichi Sasa}
\address{Department of Physics, Kyoto University, Kyoto 606-8502, Japan}
\eads{\mailto{ueda@ton.scphys.kyoto-u.ac.jp}, \mailto{sasa@scphys.kyoto-u.ac.jp}}

\begin{abstract}
We study $p$-spin glass models on regular random graphs.
By analyzing the Franz-Parisi potential with a two-body cavity field approximation under the replica symmetric ansatz, we obtain a good approximation of the 1RSB transition temperature for $p=3$.
Our calculation method is much easier than the 1RSB cavity method because the result is obtained by solving self-consistent equations with Newton's method.
\end{abstract}


\maketitle

\section{Introduction}
\label{sec:Intro}

Spin glass models on random graphs have been extensively studied in statistical mechanics of random systems \cite{ViaBra1985, MezPar1987, WonShe1988, MezPar2001, FLRZ2001, FraLeo2003, MonRic2003}.
There are two reasons for this.
The first reason comes from a motivation for describing thermodynamic phases of models with finite connectivity which might have a different nature from the infinite-range model.
Models on random graphs may form the simplest class for this motivation, while the final goal is to understand the nature of finite-dimensional models \cite{BinYou1986}.
On the other hand, apart from spin glass materials in nature, it has been found that spin glass models on random graphs frequently appear in combinatorial optimization problems \cite{MZKST1999, RWZ2001, WeiHar2001, MPWZ2002}.
Many constraint satisfaction problems can be mapped to problems concerning the ground states of spin glass models \cite{MooMer}, and the solutions space for random instances is characterized by their complexity that is calculated by a theoretical technique developed in studies on spin glasses. 
This provides the second reason for intensive studies on spin glass models on random graphs. 


The first step of studying models is to determine transition points.
As an example, let us consider $p$-body interaction spin glass models on random graphs \cite{MezMon}.
For the model with $p=2$, a full replica symmetry breaking (FRSB) occurs at a transition temperature whose value is determined by the condition that the replica symmetric (RS) solution becomes unstable.
When $p\geq 3$, a one-step RSB (1RSB) occurs, but the transition is not detected by simply searching the instability point of the RS solution because of the discontinuous nature of the transition.
In order to determine the transition temperature, we need another method different from analyzing the free energy under the RS ansatz.


Up to the present, the 1RSB transition temperature of models on random graphs has often been studied by using the replica method with the 1RSB ansatz \cite{Mona1998}, a variational approximation of the replica method \cite{BMW2000}, the 1RSB cavity method extending the Bethe-Peierls approximation to the 1RSB phase \cite{MezPar2001}, and the replica method with finite replica number and the RS ansatz \cite{NakHuk2008, NakHuk2009}.
In particular, in the 1RSB cavity method, which might be the most standard method for the calculation of the 1RSB transition temperature, one needs to solve self-consistent equations for functionals of the probability distributions of cavity fields in general.
These self-consistent equations can be solved numerically by using the algorithm called population dynamics. 
It should be noted that the 1RSB transition is characterized by the divergence of point-to-set correlation length within the 1RSB cavity method \cite{MezMon2006}.
In the replica method for models on random graphs, the free energy is calculated by using the function order parameter.
The method using finite replica number \cite{NakHuk2009} assumes a property of the cumulant generating function of the free energy, which turns out to be equivalent to the replica method with the factorized ansatz \cite{FLRZ2001}.


As a different approach to determine the 1RSB transition temperature, the method using the Franz-Parisi potential \cite{FP1995} has been proposed.
The Franz-Parisi potential is an effective potential of overlap between two replicas and we can detect the transition by investigating a qualitative change of the potential.
An explicit form of the Franz-Parisi potential has been derived theoretically for the infinite-range spin glass models.
The advantage of the method using the Franz-Parisi potential is that we can calculate the transition temperature under only the RS ansatz.
In fact, in the infinite-range model, the transition temperature calculated by using this potential with the RS ansatz agrees with the result by the standard 1RSB ansatz.
If this method can apply to models on random graphs, we can determine the transition temperature with easier calculation than the 1RSB methods.
For models on random graphs, since the computational cost for the methods with the RS ansatz is significantly lower than that for the methods with the 1RSB ansatz, it is a major advantage if applicable.
However, the Franz-Parisi potential has not been employed for calculating the transition temperature of models on random graphs, because the 1RSB phase is characterized by a nontrivial distribution of cavity fields.
It should be noted that a computational method of the Franz-Parisi potential of diluted disordered models on random graphs was proposed by generalizing the cavity method \cite{ZdeKrz2010}.


In this paper, we calculate the transition temperature of models on regular random graphs by analyzing the Franz-Parisi potential under the RS ansatz.
By comparing our results with those obtained by the 1RSB methods, we conclude that our method is useful for obtaining a good approximation of the transition temperature with considerably lower computational cost.


This paper is organized as follows.
In section \ref{sec:Preliminaries}, we present the models we study, define the Franz-Parisi potential and its Legendre transform, and explicitly express a condition that determines the transition point.
We also express this Legendre transform by using the RS cavity method.
In section \ref{sec:Results}, we make an assumption in order to solve the problem in terms of a few self-consistent equations, and as a result, we obtain the transition temperature.
In section \ref{sec:Relation}, we remark difference between our method and the standard 1RSB methods.
Finally, in section \ref{sec:Concluding remarks}, we discuss the validity of the assumption we made.
In Appendicies, we review the analysis on the infinite-range model for reference of our method and collect details of our calculation.

\section{Preliminaries}
\label{sec:Preliminaries}
We consider the $\pm J$-type $p$-spin glass models \cite{RWZ2001, MRZ2003, CDMM2003} on a $C$-regular random graph $G$ \cite{MezMon} of size $N$, where $C$ represents connectivity.
The spin variable $\sigma_i \in \left\{ -1, 1 \right\}$ is defined on each site $i\in G$ and we write $\textrm{\boldmath $\sigma$} \equiv \left\{ \sigma_i \right\}_{i=1}^N$ collectively.
The Hamiltonian is given by
\begin{eqnarray}
 H_0 (\textrm{\boldmath $\sigma$}) &=& - \sum_{i_1< \cdots< i_p} g_{i_1, \cdots, i_p} J_{i_1, \cdots, i_p} \sigma_{i_1}\cdots \sigma_{i_p}.
\end{eqnarray}
$g_{i_1, \cdots, i_p}$ is a random variable taking a value $0$ or $1$ which is determined according to a probability distribution
\begin{eqnarray}
 P\left( \left\{ g_{i_1, \cdots, i_p} \right\} \right) &=& \prod_{i=1}^N \delta \left( \sum_{i_1< \cdots< i_{p-1}}g_{i_1, \cdots, i_{p-1}, i} - C \right).
\end{eqnarray}
$J_{i_1, \cdots, i_p}$ is a random interaction variable obeying a probability distribution
\begin{eqnarray}
 P(J) &=& \frac{1}{2} \left[ \delta_{J, 1} + \delta_{J, -1} \right].
\end{eqnarray}
We consider the case where $g_{i_1, \cdots, i_p}$ and $J_{i_1, \cdots, i_p}$ are symmetric under index permutations.
It has been known that the low temperature phase of this model is reasonably described with the 1RSB solution \cite{FLRZ2001, FMRWZ2001, MonRic2004, NakHuk2009}.
It is also known that a problem computing the ground state energy of this Hamiltonian can be mapped to a NP-hard problem, MAX-$p$-XORSAT \cite{MooMer}.

We consider an effective potential of overlap between two replicas, the Franz-Parisi potential \cite{FP1995}.
The Franz-Parisi potential was originally introduced for the infinite-range model in order to detect the appearance of metastable states as a local minimum of the potential.
In this calculation, the 1RSB transition point is determined as the temperature at which a nontrivial local minimum value of the potential is equal to the trivial local minimum value.
In other words, at the transition temperature, the overlap between two replicas takes a nontrivial value in the thermodynamic limit.
We expect that this picture can be extended to spin glass models on random graphs.
We also introduce the Legendre transform of the Franz-Parisi potential \cite{FP1997}.
To be specific, we first define the Hamiltonian with an external field directing $\textrm{\boldmath $s$}$
\begin{eqnarray}
 H_{h_\mathrm{ext}}(\textrm{\boldmath $\sigma$}; \textrm{\boldmath $s$}) &=& - \sum_{i_1< \cdots< i_p} g_{i_1, \cdots, i_p} J_{i_1, \cdots, i_p} \sigma_{i_1}\cdots \sigma_{i_p} - h_\mathrm{ext} \sum_{i=1}^N s_i \sigma_i.
\end{eqnarray}
The free energy of this system with inverse temperature $\beta$ and fixed $\textrm{\boldmath $s$}$ is given by $-\beta^{-1} \log\left( \sum_{\textrm{\boldmath $\sigma$}} e^{-\beta H_{h_\mathrm{ext}}(\textrm{\boldmath $\sigma$}; \textrm{\boldmath $s$})} \right)$.
We then assume that $\textrm{\boldmath $s$}$ obeys the canonical distribution with the Hamiltonian $H_0$ and inverse temperature $\beta^\prime$.
By taking the average of the free energy with respect to $\textrm{\boldmath $s$}$, $g_{i_1, \cdots, i_p}$ and $J_{i_1, \cdots, i_p}$, we obtain the Legendre transform of the Franz-Parisi potential
\begin{eqnarray}
 - \beta \mathcal{G}(\beta, \beta^\prime, h_\mathrm{ext}) &\equiv& \mathbb{E}_g \mathbb{E}_J \left[ \frac{1}{Z_0^\prime} \sum_{\textrm{\boldmath $s$}} e^{-\beta^\prime H_0(\textrm{\boldmath $s$})} \log\left( \sum_{\textrm{\boldmath $\sigma$}} e^{-\beta H_{h_\mathrm{ext}}(\textrm{\boldmath $\sigma$}; \textrm{\boldmath $s$})} \right) \right],
 \label{eq:FP_free_energy}
\end{eqnarray}
where $Z_0^\prime \equiv \sum_{\textrm{\boldmath $s$}} e^{-\beta^\prime H_0(\textrm{\boldmath $s$})}$, and $\mathbb{E}_g$ and $\mathbb{E}_J$ represent the expected value with respect to $g$ and $J$, respectively.
$-\beta \mathcal{G}$ corresponds to the cumulant generating function of the overlap $N^{-1}\sum_{i=1}^N s_i \sigma_i$.
In the infinite-range model, we could calculate the 1RSB transition temperature only with the RS ansatz.
(The details are given in \ref{meanfield}.)
Based on this achievement, we attempt to calculate $-\beta \mathcal{G}(\beta, \beta^\prime, h_\mathrm{ext})$ by using the replica method for the models on random graphs.

In order to obtain the transition temperature, we have only to calculate the cumulant generating function $-\beta \mathcal{G}$ for the case $\beta^\prime=\beta$ and $h_\mathrm{ext}=+0$.
Due to a property of the Legendre transformation, considering the case $h_\mathrm{ext}=+0$ corresponds to probing only the minima of the Franz-Parisi potential.
Under the RS ansatz, self-consistent equations in the replica method have a nontrivial solution as well as the trivial (high temperature) solution.
We denote the cumulant generating functions calculated from the nontrivial solution and the trivial solution by $-\beta \mathcal{G}_\mathrm{SG}$ and $-\beta \mathcal{G}_\mathrm{para}$, respectively.
We assume that the nontrivial RS solution is stable above the 1RSB transition temperature $T_\mathrm{K}$.
It should be noted that this trivial and nontrivial solution correspond to the global minima and local minima of the Franz-Parisi potential, respectively.
The transition temperature $T_\mathrm{K}$ is determined by the condition
\begin{eqnarray}
 -\frac{1}{N}\beta \mathcal{G}_\mathrm{SG}(\beta, \beta, +0) &=& -\frac{1}{N}\beta \mathcal{G}_\mathrm{para}(\beta).
 \label{eq:FP_transition}
\end{eqnarray}
Below we assume $p=3$.
It is straightforward to extend our method to $p\geq 4$ case.

We calculate the cumulant generating function (\ref{eq:FP_free_energy}) under the RS ansatz.
By using the replica method \cite{FP1995}, we introduce two replica numbers $n$ and $R$.
We first define a free energy of the $R$-replicated system as
\begin{eqnarray}
 \fl -\beta g_R &\equiv& \frac{1}{N}\mathbb{E}\left[ \log\left( \sum_{\textrm{\boldmath $s$}} \sum_{\left\{ \textrm{\boldmath $\sigma$}^{(r)} \right\}} e^{-\beta^\prime H_0(\textrm{\boldmath $s$}) - \beta \sum_{r=1}^{R-1} H_0(\textrm{\boldmath $\sigma$}^{(r)}) + \beta h_\mathrm{ext} \sum_{r=1}^{R-1} \textrm{\boldmath $\sigma$}^{(r)}\cdot \textrm{\boldmath $s$}} \right) \right].
\end{eqnarray}
The cumulant generating function $-\beta \mathcal{G}$ is then expressed as
\begin{eqnarray}
 \fl && -\frac{1}{N}\beta \mathcal{G}(\beta, \beta^\prime, h_\mathrm{ext}) \nonumber \\
 \fl &=& \frac{1}{N} \lim_{n\rightarrow 0} \lim_{R\rightarrow 1} \frac{1}{n} \mathbb{E} \left[ \frac{\partial }{\partial R} \left\{ \sum_{\textrm{\boldmath $s$}} e^{-\beta^\prime H_0(\textrm{\boldmath $s$})} \left( \sum_{\textrm{\boldmath $\sigma$}} e^{-\beta H_0(\textrm{\boldmath $\sigma$}) + \beta h_\mathrm{ext} \textrm{\boldmath $\sigma$}\cdot \textrm{\boldmath $s$}} \right)^{R-1} \right\}^n \right] \nonumber \\
 \fl &=& \frac{1}{N} \lim_{n\rightarrow 0} \lim_{R\rightarrow 1} \frac{\mathbb{E}\left[ \left\{ \sum_{\textrm{\boldmath $s$}} e^{-\beta^\prime H_0(\textrm{\boldmath $s$})} \left( \sum_{\textrm{\boldmath $\sigma$}} e^{-\beta H_0(\textrm{\boldmath $\sigma$}) + \beta h_\mathrm{ext} \textrm{\boldmath $\sigma$}\cdot \textrm{\boldmath $s$}} \right)^{R-1} \right\}^n \right] - \mathbb{E}\left[ \left\{ \sum_{\textrm{\boldmath $s$}} e^{-\beta^\prime H_0(\textrm{\boldmath $s$})} \right\}^n \right]}{n(R-1)} \nonumber \\
 \fl &=& \lim_{R \rightarrow 1} \frac{-\beta g_R - \left( -\beta g_1 \right)}{R-1}.
\end{eqnarray}
Hereafter, we set $\textrm{\boldmath $\sigma$}^{(0)} \equiv \textrm{\boldmath $s$}$.
We can use the cavity method for calculating $-\beta g_R$.
We denote $R$ spins collectively by $\tau \equiv \left(\sigma^{(0)}, \sigma^{(1)}, \cdots, \sigma^{(R-1)} \right)$ for a site in $G$.
By applying the RS cavity method \cite{MezMon}, $-\beta g_R$ is expressed in terms of $k$-body cavity fields $\mathfrak{h} \equiv \left[ \left\{ h^{(r_1, \cdots, r_k)} \right\}_{r_1< \cdots< r_k} \right]_{1\leq k\leq R}$.
We also define $\mathfrak{h}_\mathrm{ext}$ as the quantity which takes the value $h_\mathrm{ext}$ for the $(0, r)$ components with $1 \leq r \leq R-1$ and $0$ otherwise, for convenience.
The free energy is expressed as
\begin{eqnarray}
 -\beta g_R &=& \int \prod_{b=1}^C d\hat{P}\left( \hat{\mathfrak{h}}_b \right) \log w\left( \mathfrak{h}_\mathrm{ext} + \sum_{b=1}^C \hat{\mathfrak{h}}_b \right) \nonumber \\
 && \qquad - C \int d\hat{P}\left( \hat{\mathfrak{h}} \right) \int dP\left( \mathfrak{h} \right) \log w\left( \mathfrak{h} + \hat{\mathfrak{h}} \right) \nonumber \\
 && \qquad + \frac{C}{3} \int \prod_{j=1}^3 dP\left( \mathfrak{h}_j \right) \mathbb{E}_J \left[ \log I\left( \left\{ \mathfrak{h}_j \right\} \right) \right],
 \label{eq:g_R}
\end{eqnarray}
where we set $\beta_0^{(0)}\equiv \beta^\prime$ and $\beta_0^{(r)}\equiv \beta \: (1 \leq r \leq R-1)$, and $w\left( \mathfrak{h} \right)$ and $I\left( \left\{ \mathfrak{h}_j \right\} \right)$ in (\ref{eq:g_R}) are given by
\begin{eqnarray}
 w\left( \mathfrak{h} \right) &\equiv& \sum_\tau e^{\sum_{k=1}^R \sum_{r_1< \cdots< r_k} \beta h^{(r_1, \cdots, r_k)} \sigma^{(r_1)}\cdots \sigma^{(r_k)}},
 \label{eq:def_w} \\
 I\left( \left\{ \mathfrak{h}_j \right\} \right) &\equiv& \sum_{\tau_{(1)}, \tau_{(2)}, \tau_{(3)}} e^{\sum_{j=1}^3 \sum_{k=1}^R \sum_{r_1< \cdots< r_k} \beta h_j^{(r_1, \cdots, r_k)} \sigma_{(j)}^{(r_1)}\cdots \sigma_{(j)}^{(r_k)}} \nonumber \\
 && \qquad \times e^{J\sum_{r=0}^{R-1}\beta_0^{(r)}\sigma_{(1)}^{(r)}\sigma_{(2)}^{(r)}\sigma_{(3)}^{(r)}}.
 \label{eq:def_I}
\end{eqnarray}
The probability distributions $P$ and $\hat{P}$ satisfy the following self-consistent equations called density evolution equations:
\begin{eqnarray}
 \fl \hat{P}\left( \hat{\mathfrak{h}} \right) &=& \int \prod_{j=1}^2 dP\left( \mathfrak{h}_j \right) \mathbb{E}_J \left[\prod_{k=1}^R \prod_{r_1< \cdots< r_k} \delta\left( \hat{h}^{(r_1, \cdots, r_k)} - \frac{1}{2^R \beta} \log\left[\frac{A_1^{(r_1, \cdots, r_k)}}{A_{-1}^{(r_1, \cdots, r_k)}} \right] \right) \right],
 \label{eq:DEeq_ph} \\
 \fl P\left( \mathfrak{h} \right) &=& \int \prod_{b=1}^{C-1} d\hat{P}\left( \hat{\mathfrak{h}}_b \right) \left\{ \prod_{k\neq 2} \prod_{r_1< \cdots< r_k} \delta\left( h^{(r_1, \cdots, r_k)} - \sum_{b=1}^{C-1} \hat{h}_b^{(r_1, \cdots, r_k)} \right) \right\} \nonumber \\
 \fl && \qquad \times \left\{ \prod_{r_1< r_2}^{r_1, r_2 \neq 0} \delta\left( h^{(r_1, r_2)} - \sum_{b=1}^{C-1} \hat{h}_b^{(r_1, r_2)} \right) \right\} \nonumber \\
 \fl && \qquad \times \left\{ \prod_{r=1}^{R-1} \delta\left( h^{(0, r)} - h_\mathrm{ext} - \sum_{b=1}^{C-1} \hat{h}_b^{(0, r)} \right) \right\},
 \label{eq:DEeq_p}
\end{eqnarray}
where $A_u^{(r_1, \cdots, r_k)}$ in (\ref{eq:DEeq_ph}) is given by
\begin{eqnarray}
 A_u^{(r_1, \cdots, r_k)} &\equiv& \prod_{\tau: \sigma^{(r_1)}\cdots \sigma^{(r_k)}=u} \sum_{\tau_{(1)}, \tau_{(2)}}e^{\sum_{j=1}^2 \sum_{k^\prime=1}^R \sum_{r^\prime_1< \cdots< r^\prime_{k^\prime}} \beta h_j^{(r^\prime_1, \cdots, r^\prime_{k^\prime})} \sigma_{(j)}^{(r^\prime_1)}\cdots \sigma_{(j)}^{(r^\prime_{k^\prime})}} \nonumber \\
 && \qquad \times e^{J\sum_{r=0}^{R-1}\beta_0^{(r)}\sigma_{(1)}^{(r)}\sigma_{(2)}^{(r)}\sigma^{(r)}}.
 \label{eq:qA}
\end{eqnarray}
See reference \cite{MezMon} for details of the derivation.
We remark that these results can be derived from only the replica method by using order parameters in reference \cite{MonaZecc1997}.

For a specific natural number $R$, we can numerically calculate $-\beta g_R$ by solving the density evolution equations with a polulation dynamics algorithm \cite{MezMon}.
However it is difficult to obtain $-\beta \mathcal{G}$ on the basis of the numerical method, because we need to perform the analytic continuation of $-\beta g_R$ to real number $R$.
We attempt to derive $-\beta g_R$ without numerical evaluation.

\section{Results}
\label{sec:Results}
We assume a solution of the form
\begin{eqnarray}
 P\left( \mathfrak{h} \right) &=& \left\{ \prod_{k\neq 2} \prod_{r_1< \cdots< r_k} \delta\left( h^{(r_1, \cdots, r_k)} \right) \right\} \left\{ \prod_{r_1< r_2}^{r_1, r_2 \neq 0} \delta\left( h^{(r_1, r_2)} - h_* \right) \right\} \nonumber \\
 && \qquad \times \left\{ \prod_{r=1}^{R-1} \delta\left( h^{(0, r)} - h_{\mathrm{s}*} \right) \right\},
 \label{eq:2sol_p} \\
 \hat{P}\left( \hat{\mathfrak{h}} \right) &=& \left\{ \prod_{k\neq 2} \prod_{r_1< \cdots< r_k} \delta\left( \hat{h}^{(r_1, \cdots, r_k)} \right) \right\} \left\{ \prod_{r_1< r_2}^{r_1, r_2 \neq 0} \delta\left( \hat{h}^{(r_1, r_2)} - \hat{h}_* \right) \right\} \nonumber \\
 && \qquad \times \left\{ \prod_{r=1}^{R-1} \delta\left( \hat{h}^{(0, r)} - \hat{h}_{\mathrm{s}*} \right) \right\},
 \label{eq:2sol_ph}
\end{eqnarray}
which is exact for the infinite-range model (see \ref{meanfield}).
We will discuss the existence of this solution later.
Under this assumption, (\ref{eq:qA}) is calculated as 
\begin{eqnarray}
 \fl A_u^{(r_1, \cdots, r_k)} &=& \prod_{\tau: \sigma^{(r_1)}\cdots \sigma^{(r_k)}=u} \sum_{\tau_{(1)}, \tau_{(2)}}e^{\sum_{j=1}^2 \left[ \sum_{r^\prime_1< r^\prime_2}^{r^\prime_1, r^\prime_2 \neq 0} \beta h_* \sigma_{(j)}^{(r^\prime_1)} \sigma_{(j)}^{(r^\prime_2)} + \sum_{r=1}^{R-1}\beta h_{\mathrm{s}*}\sigma_{(j)}^{(0)}\sigma_{(j)}^{(r)} \right]} \nonumber \\
 \fl && \qquad \times e^{J\sum_{r=0}^{R-1}\beta_0^{(r)}\sigma_{(1)}^{(r)}\sigma_{(2)}^{(r)}\sigma^{(r)}} \nonumber \\
 \fl &=& \prod_{\tau: \sigma^{(r_1)}\cdots \sigma^{(r_k)}=u} 2^R e^{-\beta h_* (R-1)} \left\{ \prod_{r=0}^{R-1} \cosh\left( \beta_0^{(r)} J \right) \right\} \nonumber \\
 \fl && \qquad \times \int \prod_{j=1}^2 Dz_j \left\{ \prod_{j=1}^2 \cosh^{R-1} \left( z_j\sqrt{\beta h_*} + \beta h_{\mathrm{s}*} \right) \right\} \nonumber \\
 \fl && \qquad \times \sum_{\nu=\pm 1} \left\{ 1+ \nu\sigma^{(0)}\tanh\left( \beta^\prime J \right) \right\} \prod_{r=1}^{R-1} \left\{ 1+ \nu\sigma^{(r)}\mathrm{sgn}(J)\tanh\Theta_2 \right\},
 \label{eq:A_result}
\end{eqnarray}
where we have defined $Dz \equiv \frac{dz}{\sqrt{2\pi}} e^{-\frac{1}{2} z^2}$ and
\begin{eqnarray}
 \fl \Theta_l(z_1, \cdots, z_l; h_*, h_{\mathrm{s}*}) \equiv \tanh^{-1} \left[ \tanh\left( \beta \left| J \right| \right) \prod_{j=1}^l \tanh \left( z_j\sqrt{\beta h_*} + \beta h_{\mathrm{s}*} \right) \right].
\end{eqnarray}
(See \ref{cal}.)
By noting that (\ref{eq:A_result}) is independent of the sign of $J$, we find that the density evolution equations (\ref{eq:DEeq_ph}) and (\ref{eq:DEeq_p}) provide the following self-consistent equations for the two-body cavity fields:
\begin{eqnarray}
 h_* &=& (C-1)\hat{h}_*, \\
 h_{\mathrm{s}*} &=& h_\mathrm{ext} + (C-1)\hat{h}_{\mathrm{s}*}, \\
 \hat{h}_* &=& \frac{1}{2^R \beta} \sum_\tau \sigma^{(r_1)}\sigma^{(r_2)} \log \int \prod_{j=1}^2 Dz_j \left\{ \prod_{j=1}^2 \cosh^{R-1} \left( z_j\sqrt{\beta h_*} + \beta h_{\mathrm{s}*} \right) \right\} \nonumber \\
 && \times \sum_{\nu=\pm 1} \left\{ 1+\nu\sigma^{(0)} \tanh\left( \beta^\prime \left| J \right| \right) \right\} \prod_{r=1}^{R-1} \left\{ 1+\nu\sigma^{(r)} \tanh\Theta_2 \right\},
 \label{eq:consistent3} \\
 \hat{h}_{\mathrm{s}*} &=& \frac{1}{2^R \beta} \sum_\tau \sigma^{(0)}\sigma^{(r^\prime)} \log \int \prod_{j=1}^2 Dz_j \left\{ \prod_{j=1}^2 \cosh^{R-1} \left( z_j\sqrt{\beta h_*} + \beta h_{\mathrm{s}*} \right) \right\} \nonumber \\
 && \times \sum_{\nu=\pm 1} \left\{ 1+\nu\sigma^{(0)} \tanh\left( \beta^\prime \left| J \right| \right) \right\} \prod_{r=1}^{R-1} \left\{ 1+\nu\sigma^{(r)} \tanh\Theta_2 \right\},
 \label{eq:consistent4}
\end{eqnarray}
where $r_1, r_2 \neq 0$ and $r^\prime \neq 0$ in (\ref{eq:consistent3}) and (\ref{eq:consistent4}), respectively.
The quantity $w\left( \mathfrak{h} \right)$ in (\ref{eq:def_w}) is also calculated as
\begin{eqnarray}
 w\left( \mathfrak{h} \right) &=& \sum_\tau e^{\sum_{r=1}^{R-1}\beta h_{\mathrm{s}*}\sigma^{(0)}\sigma^{(r)} + \sum_{r_1< r_2}^{r_1, r_2 \neq 0} \beta h_* \sigma^{(r_1)}\sigma^{(r_2)}} \nonumber \\
 &=& 2^R e^{- \frac{1}{2}\beta h_* (R-1)} \int Dz \cosh^{R-1} \left( z\sqrt{\beta h_*} + \beta h_{\mathrm{s}*} \right).
\end{eqnarray}
(Details are presented in \ref{cal}.)
Furthermore, in the manner similar to $A_u^{(r_1, \cdots, r_k)}$, $I\left( \left\{ \mathfrak{h}_j \right\} \right)$ in (\ref{eq:def_I}) is calculated as
\begin{eqnarray}
 \fl I\left( \left\{ \mathfrak{h}_j \right\} \right) &=& \sum_{\tau_{(1)}, \tau_{(2)}, \tau_{(3)}} e^{\sum_{j=1}^3 \left[ \sum_{r_1< r_2}^{r_1, r_2 \neq 0} \beta h_* \sigma_{(j)}^{(r_1)} \sigma_{(j)}^{(r_2)} + \sum_{r=1}^{R-1} \beta h_{\mathrm{s}*} \sigma_{(j)}^{(0)} \sigma_{(j)}^{(r)} \right]} e^{J\sum_{r=0}^{R-1}\beta_0^{(r)}\sigma_{(1)}^{(r)}\sigma_{(2)}^{(r)}\sigma_{(3)}^{(r)}} \nonumber \\
 \fl &=& 2^{3R} e^{-\frac{3}{2}\beta h_* (R-1)} \cosh\left( \beta^\prime J \right) \cosh^{R-1}\left( \beta J \right) \nonumber \\
 \fl && \qquad \times \frac{1}{2} \int \prod_{j=1}^3 Dz_j \left\{ \prod_{j=1}^3 \cosh^{R-1} \left( z_j\sqrt{\beta h_*} + \beta h_{\mathrm{s}*} \right) \right\} \nonumber \\
 \fl && \qquad \times \sum_{\nu=\pm 1} \left\{ 1+\nu\tanh\left( \beta^\prime J \right) \right\} \left\{ 1+ \nu\mathrm{sgn}(J)\tanh\Theta_3 \right\}^{R-1},
\end{eqnarray}
which is independent of the sign of $J$ too.
By using these quantities, the free energy $g_R$ is expressed as
\begin{eqnarray}
 \fl -\beta g_R &=& R\log 2 + \frac{C}{3} \log\left[ \cosh\left( \beta^\prime \left| J \right| \right) \right] + (R-1) \frac{C}{3} \log\left[ \cosh\left( \beta \left| J \right| \right) \right] \nonumber \\
 \fl && + \log \int Dz \cosh^{R-1} \left( z\sqrt{\beta C \hat{h}_*} + \beta h_\mathrm{ext} + \beta C \hat{h}_{\mathrm{s}*} \right) \nonumber \\
 \fl && - C \log \int Dz \cosh^{R-1} \left( z\sqrt{\beta \left( h_* + \hat{h}_* \right)} + \beta \left( h_{\mathrm{s}*} + \hat{h}_{\mathrm{s}*} \right) \right) \nonumber \\
 \fl && + \frac{C}{3} \log \frac{1}{2}\int \prod_{j=1}^3 Dz_j \left\{ \prod_{j=1}^3 \cosh^{R-1} \left( z_j\sqrt{\beta h_*} + \beta h_{\mathrm{s}*} \right) \right\} \nonumber \\
 \fl && \qquad \times \sum_{\nu=\pm 1} \left\{ 1+ \nu\tanh\left( \beta^\prime \left| J \right| \right) \right\} \left\{ 1+ \nu\tanh\Theta_3 \right\}^{R-1}.
\end{eqnarray}
In particular, we obtain
\begin{eqnarray}
 -\beta g_1 = \log 2 + \frac{C}{3} \mathbb{E}_J\left[ \log \cosh\left( \beta^\prime J \right) \right].
\end{eqnarray}

Now we take the sum over $\tau$ in (\ref{eq:consistent3}) and (\ref{eq:consistent4}).
For later convenience, we define
\begin{eqnarray}
 \fl B(d_1, d_2) &\equiv& \log \int \prod_{j=1}^2 Dz_j \left\{ \prod_{j=1}^2 \cosh^{d_1+d_2} \left( z_j\sqrt{\beta h_*} + \beta h_{\mathrm{s}*} \right) \right\} \nonumber \\
 \fl && \qquad \times \sum_{\nu=\pm 1} \left\{ 1+ \nu\tanh\left( \beta^\prime \left| J \right| \right) \right\} \left\{ 1+ \nu\tanh\Theta_2 \right\}^{d_1} \left\{ 1- \nu\tanh\Theta_2 \right\}^{d_2}.
\end{eqnarray}
We also define the binomial coefficient $\binom{n}{k} \equiv \Gamma(n+1) \Gamma(k+1)^{-1} \Gamma(n-k+1)^{-1}$.
We then obtain
\begin{eqnarray}
 \hat{h}_{\mathrm{s}*} &=& \frac{1}{2^{R-1} \beta} \left( \Phi_{1,1} - \Phi_{1,2} \right), \\
 \hat{h}_* &=& \frac{1}{2^{R-1} \beta} \left( \Phi_{2,1} + \Phi_{2,2} - \Phi_{2,3} - \Phi_{2,4} \right),
\end{eqnarray}
where
\begin{eqnarray}
 \Phi_{1,1} &\equiv& \sum_{d=0}^{R-2} \binom{R-2}{d} B(R-1-d, d), \\
 \Phi_{1,2} &\equiv& \sum_{d=0}^{R-2} \binom{R-2}{d} B(d, R-1-d), \\
 \Phi_{2,1} &\equiv& \sum_{d=0}^{R-3} \binom{R-3}{d} B(R-1-d, d), \\
 \Phi_{2,2} &\equiv& \sum_{d=0}^{R-3} \binom{R-3}{d} B(d, R-1-d), \\
 \Phi_{2,3} &\equiv& \sum_{d=0}^{R-3} \binom{R-3}{d} B(R-2-d, d+1), \\
 \Phi_{2,4} &\equiv& \sum_{d=0}^{R-3} \binom{R-3}{d} B(d+1, R-2-d).
\end{eqnarray}

We consider the limit $R \rightarrow 1$ by analytic continuation of a function $\phi_f(n) \equiv \sum_{k=0}^n \binom{n}{k} f(k)$ with arbitrary $f(k)$:
\begin{eqnarray}
 \lim_{n\longrightarrow -1} \phi_f(n) &=& \left. \frac{\partial }{\partial x} \sum_{m=0}^\infty (-1)^m x^{f(m)} \right|_{x=1},
 \label{eq:identity1} \\
 \lim_{n\longrightarrow -2} \phi_f(n) &=& \left. \frac{\partial }{\partial x} \sum_{m=0}^\infty (-1)^m (m+1) x^{f(m)} \right|_{x=1}.
 \label{eq:identity2}
\end{eqnarray}
(The proof is given in \ref{AC}.)
Using these identities, we obtain the following self-consistent equations in the limit $R \rightarrow 1$:
\begin{eqnarray}
 h_{*(0)} &=& (C-1)\hat{h}_{*(0)},
 \label{eq:final_sc1} \\
 h_{\mathrm{s}*(0)} &=& h_\mathrm{ext} + (C-1)\hat{h}_{\mathrm{s}*(0)},
 \label{eq:final_sc2} \\
 \hat{h}_{\mathrm{s}*(0)} &=& \frac{1}{\beta} \left( \Phi^{(0)}_{1,1} - \Phi^{(0)}_{1,2} \right),
 \label{eq:final_sc3} \\
 \hat{h}_{*(0)} &=& \frac{1}{\beta} \left( \Phi^{(0)}_{2,1} + \Phi^{(0)}_{2,2} - \Phi^{(0)}_{2,3} - \Phi^{(0)}_{2,4} \right).
 \label{eq:final_sc4}
\end{eqnarray}
(The explicit expressions of $\Phi^{(0)}_{a,b}$ are given in \ref{AC}.)
By using $\Theta_{l(0)}(z_1, \cdots, z_l) \equiv \Theta_l(z_1, \cdots, z_l; h_{*(0)}, h_{\mathrm{s}*(0)})$, the final form of $-\beta \mathcal{G}$ is obtained as
\begin{eqnarray}
 \fl -\frac{1}{N}\beta \mathcal{G}(\beta, \beta^\prime, h_\mathrm{ext}) &=& \log 2 + \frac{C}{3} \log\left[ \cosh\left( \beta \left| J \right| \right) \right] \nonumber \\
 \fl && + \int Dz \log \cosh \left( z\sqrt{\beta C \hat{h}_{*(0)}} + \beta h_\mathrm{ext} + \beta C \hat{h}_{\mathrm{s}*(0)} \right) \nonumber \\
 \fl && - C \int Dz \log \cosh \left( z\sqrt{\beta \left( h_{*(0)} + \hat{h}_{*(0)} \right)} + \beta \left( h_{\mathrm{s}*(0)} + \hat{h}_{\mathrm{s}*(0)} \right) \right) \nonumber \\
 \fl && + C \int Dz \log \cosh \left( z\sqrt{\beta h_{*(0)} } + \beta h_{\mathrm{s}*(0)} \right) \nonumber \\
 \fl && + \frac{C}{3} \frac{1}{2}\int \prod_{j=1}^3 Dz_j \sum_{\nu=\pm 1} \left\{ 1+ \nu\tanh\left( \beta^\prime \left| J \right| \right) \right\} \nonumber \\
 \fl && \qquad \times \log \left\{ 1+ \nu\tanh\Theta_{3(0)} \right\}.
 \label{eq:final_free_energy}
\end{eqnarray}

\begin{figure}[htbp]
\begin{center}
\includegraphics{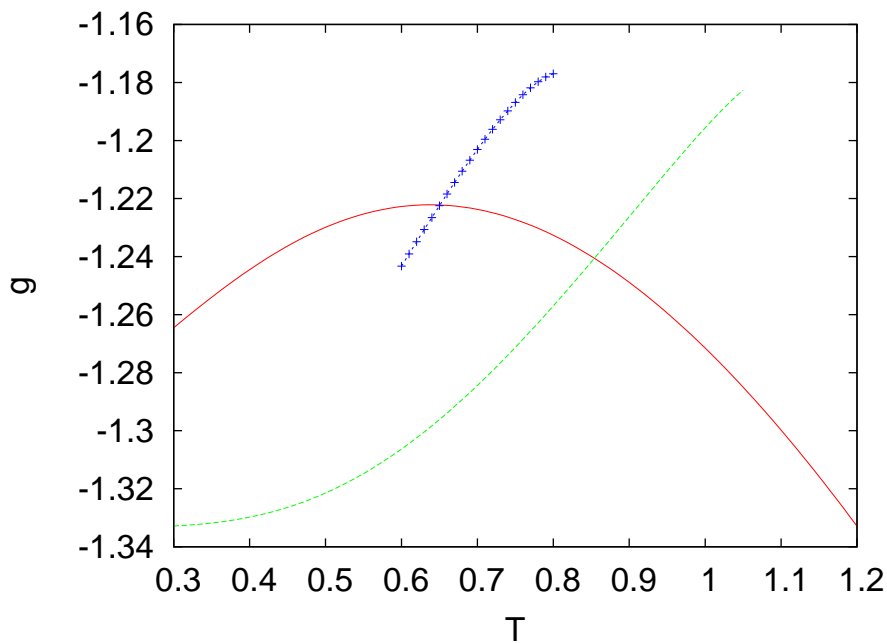}
\caption{The value of $\mathcal{G}$ for $C=4$, $\beta^\prime=\beta$ and $h_\mathrm{ext}=0$. The red curve represents the cumulant generating function calculated from the trivial solution. The blue curve represents that calculated from the nontrivial solution. The green curve represents the cumulant generating function calculated from the annealed average.}
\label{fig:free_energy_FP_3_mod_0}
\end{center}
\end{figure}

Now, we evaluate the right-hand side of (\ref{eq:final_free_energy}) numerically by solving (\ref{eq:final_sc1})-(\ref{eq:final_sc4}) with Newton's method.
The transition temperature is determined by comparing the value of the cumulant generating function calculated from a nontrivial solution, $\mathcal{G}_\mathrm{SG}$, and that calculated from the trivial solution, $\mathcal{G}_\mathrm{para}$.
The obtained results for the case $C=4$, $\beta^\prime=\beta$ and $h_\mathrm{ext}=0$ are displayed in FIG. \ref{fig:free_energy_FP_3_mod_0}.
$\mathcal{G}_\mathrm{SG}$ corresponds to the blue curve in FIG. \ref{fig:free_energy_FP_3_mod_0}.
We remark that we approximate the evaluation of infinite serieses (\ref{eq:identity1}) and (\ref{eq:identity2}) at $x=1$ as that at $x=1-\epsilon$, where $\epsilon$ is changed from $0.10$ to $0.04$, because the infinite serieses at $x=1$ are not convergent.
We checked numerically that the values of $\mathcal{G}$ for these $\epsilon$ were almost equal to each other.
$\mathcal{G}_\mathrm{para}$ is calculated from $(h_{\mathrm{s}*(0)}, h_{*(0)})=(0,0)$, which corresponds to the red curve in FIG. \ref{fig:free_energy_FP_3_mod_0}.


The static 1RSB transition point $T_\mathrm{K}$ is determined by the condition (\ref{eq:FP_transition}), which means that the blue curve crosses the red curve in FIG. \ref{fig:free_energy_FP_3_mod_0}.
The dynamical 1RSB transition point $T_\mathrm{d}$ is determined as the temperature where the nontrivial solution appears, that is, the right edge of the blue curve in FIG. \ref{fig:free_energy_FP_3_mod_0}.
In our case, we obtain $T_\mathrm{d}=0.80$ and $T_\mathrm{K}=0.65$.
For the $p=3$ spin glass model on regular random graphs with $C=4$, it has been reported $(T_\mathrm{d}, T_\mathrm{K})=(0.745, 0.660)$ \cite{FMRWZ2001} and $(T_\mathrm{d}, T_\mathrm{K})=(0.757, 0.655)$ \cite{MonRic2004} for the 1RSB cavity method, $(T_\mathrm{d}, T_\mathrm{K})=(0.752, 0.654)$ \cite{FMRWZ2001} for the replica method with the variational approximation and $T_\mathrm{K}=0.65$ \cite{NakHuk2009} for the replica method with finite replica number.
We claim from these results that our method provides a good approximation to $T_\mathrm{K}$.

It should be noted that our results are not obtained by the cumulant generating function calculated from the annealed average $\mathbb{E}\left[ \log\left( Z_0^{\prime -1} \sum_{\textrm{\boldmath $s$}} e^{-\beta^\prime H_0(\textrm{\boldmath $s$})} \sum_{\textrm{\boldmath $\sigma$}} e^{-\beta H_{h_\mathrm{ext}}(\textrm{\boldmath $\sigma$}; \textrm{\boldmath $s$})} \right) \right]$ under the RS ansatz.
This quantity is displayed as the green curve in FIG. \ref{fig:free_energy_FP_3_mod_0}.
It was shown in \cite{FP1998} that the Franz-Parisi potential calculated from the quenched average and that calculated from the annealed average are different in general.

We also give the transition temperature $T_\mathrm{K}$ for other connectivity $C$ in TABLE \ref{tab:C-T}.
These results are consistent with the previous 1RSB calculations \cite{MonRic2004, NakHuk2009}.
\begin{table}[htb]
  \begin{center}
   \caption{Connectivity dependence of transition temperature}
   \begin{tabular}{c c c}
    \hline
    Connectivity $C$ & Transition temperature $T_\mathrm{K}$ & \cite{MonRic2004}\\
    \hline
    4 & 0.651 & 0.655(5)\\
    5 & 0.848 & 0.849(5)\\
    6 & 1.00 & \\
    8 & 1.25 & 1.25(1)\\
    \hline
   \end{tabular}
   \label{tab:C-T}
 \end{center}
\end{table}
From this table we find that our method provides more accurate values as $C$ is increased.
We recall that the calculation containing only the two-body cavity fields is exact in the infinite-range model (see \ref{meanfield}).
We conjecture that the result obtained by our method approaches the exact value when $C\rightarrow \infty$.


\section{Relation with previous results}
\label{sec:Relation}
In this section, we consider the difference between our method using the Franz-Parisi potential and the previous 1RSB methods \cite{FMRWZ2001, MonRic2004, NakHuk2009}.
In our method, the transition temperature $T_\mathrm{K}$ is determined by the condition (\ref{eq:FP_transition}).
By using $-\beta L_{n, R} \equiv \lim_{N\rightarrow \infty}\frac{1}{N}\log \mathbb{E}\left[ Z_0^{nR} \right]$, the left-hand side of (\ref{eq:FP_transition}) is rewritten as 
\begin{eqnarray}
 -\frac{1}{N}\beta \mathcal{G}_\mathrm{SG}(\beta, \beta, +0) &=& \frac{1}{N} \lim_{n\rightarrow 0} \lim_{R\rightarrow 1} \frac{\mathbb{E}\left[ Z_0^{nR} \right] - \mathbb{E}\left[ Z_0^{n} \right]}{n(R-1)} \nonumber \\
 &=& \frac{1}{N} \lim_{n\rightarrow 0} \lim_{R\rightarrow 1} \frac{e^{-N\beta L_{n, R}} - e^{-N\beta L_{n, 1}}}{n(R-1)}.
 \label{eq:GFPcal1}
\end{eqnarray}
We evaluate $L_{n, R}$ under the RS ansatz.
Here we define $-\beta L_{R} \equiv \lim_{N\rightarrow \infty}\frac{1}{N}\log \mathbb{E}\left[ Z_0^{R} \right]$, which corresponds to the cumulant generating function of the free energy $\log Z_0$.
In the infinite-range model, by choosing an appropriate form of the overlap matrix as (\ref{eq:ov_FP_RS}), we obtain $L_{n, R}$ of the form
\begin{eqnarray}
 -\beta L^{(\mathrm{RS})}_{n, R} = -n\beta L^{(\mathrm{RS})}_{R}.
 \label{eq:FPass1}
\end{eqnarray}
In this case, we can rewrite the right-hand side of (\ref{eq:GFPcal1}) as
\begin{eqnarray}
 -\frac{1}{N}\beta \mathcal{G}_\mathrm{SG}(\beta, \beta, +0) &=& \lim_{R\rightarrow 1} \frac{-\beta L^{(\mathrm{RS})}_{R} - \left( -\beta L^{(\mathrm{RS})}_{1} \right)}{R-1} \nonumber \\
 &=& \left. \frac{\partial \left( -\beta L^{(\mathrm{RS})}_{R} \right)}{\partial R} \right|_{R=1}.
\end{eqnarray}
Furthermore, for the infinite-range model, the relation
\begin{eqnarray}
 -\beta L^{(\mathrm{RS})}_{1} &=& -\frac{1}{N}\beta \mathcal{G}_\mathrm{para}(\beta)
 \label{eq:FPass2}
\end{eqnarray}
holds.
By using this relation, we can rewrite the phase transition condition (\ref{eq:FP_transition}) as
\begin{eqnarray}
 0 &=& \left. \frac{\partial \left( -\beta L^{(\mathrm{RS})}_R \right)}{\partial R} \right|_{R=1} - \left( -\beta L^{(\mathrm{RS})}_1 \right) \nonumber \\
 &=& \left. \frac{\partial }{\partial R} \left[ -\frac{1}{R}\beta L^{(\mathrm{RS})}_R \right] \right|_{R=1}.
\end{eqnarray}
This condition is equivalent to the phase transition condition in the method using finite replica number \cite{NakHuk2008}.
Thus, the method using the Franz-Parisi potential and the method using finite replica number give the same transition temperature for the infinite-range model.

However, for models on random graphs, the relations (\ref{eq:FPass1}) and (\ref{eq:FPass2}) do not hold in general, as we can see from reference \cite{NakHuk2009}.
Thus, the phase transition condition calculated with the Franz-Parisi potential
\begin{eqnarray}
 0 = \lim_{R\rightarrow 1} \frac{\left.\mathbb{E}\left[ \log Z_0^R \right] \right|_\mathrm{RS} - \left.\mathbb{E}\left[ \log Z_0 \right] \right|_\mathrm{RS}}{R-1} - \left. \mathbb{E}\left[ \log Z_0 \right] \right|_\mathrm{RS}
\end{eqnarray}
is generally not equivalent to the phase transition condition calculated by using finite replica number
\begin{eqnarray}
 0 = \lim_{R\rightarrow 1} \frac{\log \left.\mathbb{E}\left[ Z_0^R \right] \right|_\mathrm{RS} - \log \left.\mathbb{E}\left[ Z_0 \right] \right|_\mathrm{RS}}{R-1} - \log \left. \mathbb{E}\left[ Z_0 \right] \right|_\mathrm{RS}.
\end{eqnarray}
For models on regular random graphs, since the method using finite replica number is equivalent to the 1RSB replica method with the factorized ansatz \cite{FLRZ2001} and the 1RSB cavity method with the homogeneous assumption \cite{MonRic2004} respectively, we conclude that our calculation method using the Franz-Parisi potential is not equivalent to these 1RSB calculations.


\section{Concluding remarks}
\label{sec:Concluding remarks}
Before concluding the paper, we present three remarks.
First, we remark that calculation containing only the two-body cavity field is not exact.
For our calculation to be exact, the $k$-body cavity fields with $k\neq 2$ should be zero in the density evolution equation (\ref{eq:DEeq_ph}) with the assumed form (\ref{eq:2sol_p}).
However, it turns out that the $k$-body cavity fields take finite vaules, because, as one example, the quantity appearing in (\ref{eq:DEeq_ph})
\begin{eqnarray}
 \fl \frac{1}{2^R \beta} \log\left[\frac{A_1^{(r_1, \cdots, r_k)}}{A_{-1}^{(r_1, \cdots, r_k)}} \right] &=& \frac{1}{2^R \beta} \sum_\tau \sigma^{(r_1)}\cdots \sigma^{(r_k)} \nonumber \\
 \fl && \qquad \times \log \int \prod_{j=1}^2 Dz_j \left\{ \prod_{j=1}^2 \cosh^{R-1} \left( z_j\sqrt{\beta h_*} + \beta h_{\mathrm{s}*} \right) \right\} \nonumber \\
 \fl && \qquad \times \sum_{\nu=\pm 1} \left\{ 1+\nu\sigma^{(0)} \tanh\left( \beta^\prime \left| J \right| \right) \right\} \prod_{r=1}^{R-1} \left\{ 1+\nu\sigma^{(r)} \tanh\Theta_2 \right\}
\end{eqnarray}
is finite for even number $k\neq 2$.
In other words, our results provide approximations.
Nevertheless, it should be noted that we obtain the transition temperature $T_\mathrm{K}$ close to those obtained by the 1RSB methods.
This fact suggests that the two-body cavity fields play an important role in describing RSB even in the models on random graphs.

Second, we conjecture that it would be difficult to extend our method to models on inhomogeneous graphs such as Erd\"{o}s-R\'{e}nyi graphs, because the assumption of homogeneous solution (\ref{eq:2sol_p}) and (\ref{eq:2sol_ph}) may not be effective in this case.
We need to develop another method so as to analyze models on inhomogeneous graphs.

We finally remark whether our method using the Franz-Parisi potential with the two-body cavity field approximation can be applied to other models on regular random graphs.
We conjecture that our method is useful for models whose fully-connected limit can be described only by the two-body cavity fields.
Therefore, for example, we expect that the Potts glass models \cite{KrzZde2008}, whose fully-connected limit is described by using only one and two-body overlaps, can be analyzed by our method.
Nevertheless, we do not know the extent of how our method is useful for other cases such as lattice glass models \cite{BM2004} that cannot be defined on fully-connected graphs.
We will study these problems elsewhere in future.

In summary, we calculated the approximation value of the 1RSB transition temperature of the $p$-spin glass models on regular random graphs by analyzing the Franz-Parisi potential with the two-body cavity field approximation under the replica symmetric ansatz.
Our results are consistent with those calculated by the 1RSB methods and the computational cost for calculating transition temperatures is substantially reduced in our method, because we can determine $T_\mathrm{K}$ only by solving the four-variable self-consistent equations (\ref{eq:final_sc1})-(\ref{eq:final_sc4}), in contrast to the 1RSB cavity method, where self-consistent equations for probability distribution of cavity field should be solved.

\ack
We thank S. Takabe and K. Hukushima for valuable discussions particularly on the expressions of cumulant generating functions.
The numerical calculations were carried out on SR16000 at YITP in Kyoto University.
This work was supported by KAKENHI Nos. 22340109 and 25103002 and by the JSPS Core-to-Core Program ``Nonequilibrium dynamics of soft matter and information''.

\appendix

\section{} \label{meanfield}
In this section, we review the calculation method of the 1RSB transition temperature by using the Franz-Parisi potential for the infinite-range spin glass model.
The Hamiltonian we study is given by
\begin{eqnarray}
 H_0(\textrm{\boldmath $\sigma$}) &=& - \sum_{i_1< \cdots< i_p} J_{i_1, \cdots, i_p} \sigma_{i_1}\cdots \sigma_{i_p},
\end{eqnarray}
where $J_{i_1, \cdots, i_p}$ is a random interaction variable obeying a Gaussian distribution
\begin{eqnarray}
  P(J_{i_1, \cdots, i_p}) &=&  \sqrt{\frac{N^{p-1}}{\pi p!}}e^{-\frac{N^{p-1}J_{i_p, \cdots, i_p}^2}{p!}}.
\end{eqnarray}
Below we assume $p=3$.
The Legendre transform of the Franz-Parisi potential is defined by
\begin{eqnarray}
 \fl -\beta g(\beta, \beta^\prime; h_\mathrm{ext}) &\equiv&  \lim_{N\rightarrow \infty}\frac{1}{N} \mathbb{E} \left[ \frac{1}{Z_0^\prime} \sum_{\textrm{\boldmath $s$}} e^{-\beta^\prime H_0(\textrm{\boldmath $s$})} \log\left( \sum_{\textrm{\boldmath $\sigma$}} e^{-\beta H_0(\textrm{\boldmath $\sigma$}) + \beta h_\mathrm{ext} \sum_{i=1}^N \sigma_is_i} \right) \right],
\end{eqnarray}
where $\mathbb{E}$ represents an average over the quenched random coupling $J_{i_1, \cdots, i_p}$.
The transition point is determined by the condition (\ref{eq:FP_transition}), that is,
\begin{eqnarray}
 -\beta g_\mathrm{SG}(\beta, \beta; +0) &=& -\beta g_\mathrm{para}(\beta).
 \label{eq:TK_mf}
\end{eqnarray}

First, we calculate the transition point by using overlaps.
$-\beta g$ is rewritten by using the replica method \cite{FP1995} as
\begin{eqnarray}
 -\beta g(\beta, \beta^\prime; h_\mathrm{ext}) &=& \lim_{N\rightarrow \infty}\frac{1}{N} \lim_{n\rightarrow 0} \lim_{R\rightarrow 1} \frac{\mathbb{E}\left[ Z_{n,R} \right] - \mathbb{E}\left[ Z_{n,1} \right]}{n(R-1)},
 \label{eq:g_rep_ov}
\end{eqnarray}
where we have defined the partition function of a replicated system as
\begin{eqnarray}
 \mathbb{E}\left[ Z_{n,R} \right] &\equiv& \mathbb{E}\left[ \sum_{\left\{ \textrm{\boldmath $s$}^{(a)} \right\}} \sum_{\left\{ \textrm{\boldmath $\sigma$}^{(a,r)} \right\}} e^{-\beta^\prime \sum_{a=1}^n H_0(\textrm{\boldmath $s$}^{(a)}) - \beta \sum_{r=1}^{R-1} \sum_{a=1}^n H_0(\textrm{\boldmath $\sigma$}^{(a,r)})} \right. \nonumber \\
 && \qquad \left. \times  e^{\beta h_\mathrm{ext} \sum_{r=1}^{R-1} \sum_{a=1}^n \textrm{\boldmath $\sigma$}^{(a,r)}\cdot\textrm{\boldmath $s$}^{(a)}} \right].
 \label{eq:ZR_def_ov}
\end{eqnarray}
By performing the configurational average $\mathbb{E}$ and introducing overlaps $q_{(a_1,r_1)(a_2,r_2)} \equiv \frac{1}{N}\textrm{\boldmath $\sigma$}^{(a_1,r_1)}\cdot\textrm{\boldmath $\sigma$}^{(a_2,r_2)}$, $\mathbb{E}\left[ Z_{n,R} \right]$ is expressed as
\begin{eqnarray}
 \fl \mathbb{E}\left[ Z_{n,R} \right] &=& \int \prod dq_{(a_1,r_1)(a_2,r_2)} \int \prod d\lambda_{(a_1,r_1)(a_2,r_2)} \nonumber \\
 \fl && \qquad \times \sum_{\left\{ \textrm{\boldmath $s$}^{(a)} \right\}} \sum_{\left\{ \textrm{\boldmath $\sigma$}^{(a,r)} \right\}} e^{N\beta h_\mathrm{ext} \sum_{r=1}^{R-1} \sum_{a=1}^n q_{(a,0)(a,r)}} \nonumber \\
 \fl && \qquad \times e^{\frac{N}{4}\sum_{r_1=0}^{R-1} \sum_{a_1=1}^n \sum_{r_2=0}^{R-1} \sum_{a_2=1}^n \beta^{(r_1)} \beta^{(r_2)} q_{(a_1,r_1)(a_2,r_2)}^p} \nonumber \\
 \fl && \qquad \times e^{-iN\sum_{a_1,a_2}\sum_{r_1,r_2} \lambda_{(a_1,r_1)(a_2,r_2)} \left( q_{(a_1,r_1)(a_2,r_2)} - \frac{1}{N}\textrm{\boldmath $\sigma$}^{(a_1,r_1)}\cdot\textrm{\boldmath $\sigma$}^{(a_2,r_2)} \right)} \nonumber \\
 \fl &\sim& e^{-N\beta g_{n,R}}, 
 \label{eq:ZR_cal_ov}
\end{eqnarray}
where we set $\textrm{\boldmath $\sigma$}^{(a,0)} \equiv \textrm{\boldmath $s$}^{(a)}$ and
\begin{eqnarray}
 \beta^{(r)} &\equiv& \left\{
 \begin{array}{ll}
 \beta^\prime &\quad (r=0) \\
 \beta &\quad (r\geq 1).
 \end{array}
 \right.
\end{eqnarray}
$-\beta g_{n,R}$ is calculated by the saddle point method.
We consider the case that $\beta^\prime=\beta$ and $h_\mathrm{ext}=0$, because only this case is needed for calculating the transition temperature (\ref{eq:TK_mf}).
Since the spin variables $\textrm{\boldmath $\sigma$}^{(a,0)}$ are no longer special in this situation, we can assume the RS ansatz in which all replica indexes $r$ are treated equally and only overlaps between the same $a$'s are nonzero as
\begin{eqnarray}
 q_{(a_1,r_1)(a_2,r_2)} &=& \delta_{a_1,a_2}\left[ \delta_{r_1,r_2} + q\left( 1-\delta_{r_1,r_2} \right) \right].
 \label{eq:ov_FP_RS}
\end{eqnarray}
The matrix $\lambda_{(a_1,r_1)(a_2,r_2)}$ is assumed to take the same form.
Under this ansatz, the saddle point equations for (\ref{eq:ZR_cal_ov}) are written as
\begin{eqnarray}
 i\lambda &=& \frac{1}{2}p\beta^2q^{p-1},
 \label{eq:speq_ov_mf1} \\
 q &=& \frac{\int \frac{dz}{\sqrt{2\pi}} e^{-\frac{1}{2}z^2} \tanh^2\left(z \sqrt{i\lambda} \right) \cosh^R\left(z \sqrt{i\lambda} \right)}{\int \frac{dz}{\sqrt{2\pi}} e^{-\frac{1}{2}z^2} \cosh^R\left(z \sqrt{i\lambda} \right)}.
 \label{eq:speq_ov_mf2}
\end{eqnarray}
By using the RS ansatz (\ref{eq:ov_FP_RS}) for (\ref{eq:ZR_cal_ov}), the quantity $g_{n,R}$ is calculated as
\begin{eqnarray}
 \fl -\beta g_{n,R} &=& \frac{1}{4}nR\beta^2 + \frac{1}{4}nR(R-1)\beta^2q^p - \frac{1}{2}nR(R-1)i\lambda q - \frac{1}{2}nRi\lambda + nR\log 2 \nonumber \\
 \fl && \qquad + n\log \int\frac{dz}{\sqrt{2\pi}} e^{-\frac{1}{2}z^2}\cosh^R\left(z \sqrt{i\lambda} \right).
 \label{eq:gnR_ov_mf}
\end{eqnarray}
By using $-\beta g_{n,1}=\frac{1}{4}n\beta^2+n\log 2$, $-\beta g$ is finally calculated from (\ref{eq:g_rep_ov}):
\begin{eqnarray}
 \fl -\beta g(\beta, \beta; +0) &=& \frac{1}{4}\beta^2 + \log 2 + \frac{1}{4}\beta^2q^p - \frac{1}{2}i\lambda q - \frac{1}{2}i\lambda \nonumber \\
 \fl && \qquad + \frac{\int \frac{dz}{\sqrt{2\pi}} e^{-\frac{1}{2}z^2} \cosh\left(z \sqrt{i\lambda} \right) \log\cosh\left(z \sqrt{i\lambda} \right)}{\int \frac{dz}{\sqrt{2\pi}} e^{-\frac{1}{2}z^2} \cosh\left(z \sqrt{i\lambda} \right)}.
\end{eqnarray}
Because the cumulant generating function calculated from the trivial solution is obtained as $-\beta g_\mathrm{para}(\beta) = \frac{1}{4}\beta^2 + \log 2$, the phase transition condition (\ref{eq:TK_mf}) is expressed as
\begin{eqnarray}
 \fl 0 &=& \frac{1}{4}\beta^2q^p - \frac{1}{2}i\lambda q - \frac{1}{2}i\lambda + \frac{\int \frac{dz}{\sqrt{2\pi}} e^{-\frac{1}{2}z^2} \cosh\left(z \sqrt{i\lambda} \right) \log\cosh\left(z \sqrt{i\lambda} \right)}{\int \frac{dz}{\sqrt{2\pi}} e^{-\frac{1}{2}z^2} \cosh\left(z \sqrt{i\lambda} \right)},
\end{eqnarray}
which is equivalent to the phase transition condition under the 1RSB ansatz \cite{MezMon}.
It should be noted that the trivial solution with $q=0$ and a nontrivial solution with non-zero $q$ to Eq. (\ref{eq:speq_ov_mf2}) with $R=1$ correspond to the global minima and local minima of the Franz-Parisi potential, respectively.

Secondly, we also calculate $-\beta g$ by using the method with Monasson's order parameters \cite{MonaZecc1997}.
In this method, we evaluate the partition function (\ref{eq:ZR_def_ov}) using Monasson's order parameters, which are generalized magnetizations to $2^{nR}-1$ directions.
$-\beta g$ is expressed as
\begin{eqnarray}
 -\beta g(\beta, \beta^\prime; h_\mathrm{ext}) &=& \lim_{n\rightarrow 0} \lim_{R \rightarrow 1} \frac{-\beta g_{n,R} - \left( -\beta g_{n,1} \right)}{n(R-1)}, 
\end{eqnarray}
where by using the notation $\vec{\tau} \equiv \left(\vec{\sigma}^{(0)}, \vec{\sigma}^{(1)}, \cdots, \vec{\sigma}^{(R-1)} \right)$ with $\vec{\sigma}^{(r)} \equiv \left( \sigma^{(1,r)}, \cdots, \sigma^{(n,r)} \right)$ and $\vec{s} \equiv \left( s^{(1)}, \cdots, s^{(n)} \right)$, $g_{n,R}$ has been defined as
\begin{eqnarray}
 -\beta g_{n,R} &=& - \sum_{\vec{\tau}} m\left( \vec{\tau} \right) \log m\left( \vec{\tau} \right) + \beta h_\mathrm{ext} \sum_{\vec{\tau}} m\left( \vec{\tau} \right) \sum_{a=1}^n \sum_{r=1}^{R-1}\sigma^{(a,0)}\sigma^{(a,r)} \nonumber \\
 && + \frac{N^2}{3!}\sum_{\vec{\tau}_{(1)}} \sum_{\vec{\tau}_{(2)}} \sum_{\vec{\tau}_{(3)}} m\left( \vec{\tau}_{(1)} \right) m\left( \vec{\tau}_{(2)} \right) m\left( \vec{\tau}_{(3)} \right) \nonumber \\
 && \qquad \times \log \mathbb{E} \left[ e^{J\sum_{a=1}^n \sum_{r=0}^{R-1} \beta^{(r)} \sigma_{(1)}^{(a,r)} \sigma_{(2)}^{(a,r)} \sigma_{(3)}^{(a,r)}} \right].
\end{eqnarray}
In particular, $g_{n,1}$ is expressed as
\begin{eqnarray}
 -\beta g_{n,1} &=& - \sum_{\vec{s}} m^{(\mathrm{ref})}\left( \vec{s} \right) \log m^{(\mathrm{ref})}\left( \vec{s} \right) \nonumber \\
 && + \frac{N^2}{3!}\sum_{\vec{s}_{(1)}}\sum_{\vec{s}_{(2)}}\sum_{\vec{s}_{(3)}}m^{(\mathrm{ref})}\left( \vec{s}_{(1)} \right)m^{(\mathrm{ref})}\left( \vec{s}_{(2)} \right)m^{(\mathrm{ref})}\left( \vec{s}_{(3)} \right) \nonumber \\
 && \qquad \times \log \mathbb{E} \left[ e^{J\sum_{a=1}^n \beta^\prime s_{(1)}^{(a)} s_{(2)}^{(a)} s_{(3)}^{(a)}} \right].
\end{eqnarray}
Monasson's order parameters $m\left( \vec{\tau} \right)$ and $m^{(\mathrm{ref})}\left( \vec{s} \right)$ satisfy the self-consistent equations
\begin{eqnarray}
 \fl 0 &=& \lambda - m\left( \vec{\tau} \right) + \beta h_\mathrm{ext} \sum_{\vec{\tau}} \sum_{a=1}^n \sum_{r=1}^{R-1}\sigma^{(a,0)}\sigma^{(a,r)} \nonumber \\
 \fl && \qquad + \frac{N^2}{2} \sum_{\vec{\tau}_{(1)}} \sum_{\vec{\tau}_{(2)}} m\left( \vec{\tau}_{(1)} \right) m\left( \vec{\tau}_{(2)} \right) \log \mathbb{E} \left[ e^{J\sum_{a=1}^n \sum_{r=0}^{R-1} \beta^{(r)} \sigma_{(1)}^{(a,r)} \sigma_{(2)}^{(a,r)} \sigma^{(a,r)}} \right], \\
 \fl 0 &=& \lambda^{(\mathrm{ref})} - m^{(\mathrm{ref})}\left( \vec{s} \right) \nonumber \\
 \fl && \qquad + \frac{N^2}{2} \sum_{\vec{s}_{(1)}} \sum_{\vec{s}_{(2)}} m^{(\mathrm{ref})}\left( \vec{s}_{(1)} \right) m^{(\mathrm{ref})}\left( \vec{s}_{(2)} \right) \log \mathbb{E} \left[ e^{J\sum_{a=1}^n \beta^\prime s_{(1)}^{(a)} s_{(2)}^{(a)} s^{(a)}} \right],
\end{eqnarray}
where $\lambda$ and $\lambda^{(\mathrm{ref})}$ are determined by the normalization condition $\sum_{\vec{\tau}}m\left( \vec{\tau} \right) = 1$ and $\sum_{\vec{s}}m^{(\mathrm{ref})}\left( \vec{s} \right) = 1$.
We consider the case $\beta^\prime=\beta$ and $h_\mathrm{ext}=0$, because only this case is needed for the calculation of the transition temperature.
Since the spin variables $\textrm{\boldmath $\sigma$}^{(a,0)}$ are no longer special in this situation in the $R$-replicated system, we assume the RS ansatz as
\begin{eqnarray}
 m\left( \vec{\tau} \right) &=& \int d\mathfrak{h} P\left( \mathfrak{h} \right) \frac{e^{\beta \sum_{k=1}^{R} \sum_{r_1< \cdots< r_k} h^{(r_1, \cdots, r_k)} \sum_{a=1}^n \sigma^{(a,r_1)}\cdots \sigma^{(a,r_k)}}}{w\left( \mathfrak{h} \right)^n},
\end{eqnarray}
where we have defined 
\begin{eqnarray}
 w\left( \mathfrak{h} \right) \equiv \sum_{\tau} e^{\beta \sum_{k=1}^{R} \sum_{r_1< \cdots< r_k} h^{(r_1, \cdots, r_k)}\sigma^{(r_1)}\cdots \sigma^{(r_k)}}.
\end{eqnarray}
Furthermore, by assuming the two-body cavity field solution
\begin{eqnarray}
 P\left( \mathfrak{h} \right) &=& \left\{ \prod_{k\neq 2} \prod_{r_1< \cdots< r_k} \delta\left( h^{(r_1, \cdots, r_k)} \right) \right\} \left\{ \prod_{r_1< r_2} \delta\left( h^{(r_1, r_2)} - \frac{3}{2}\beta D_e^{2} \right) \right\},
\end{eqnarray}
the self-consistent equation is rewritten as
\begin{eqnarray}
 \fl D_e &\equiv& \frac{\int \frac{dz}{\sqrt{2\pi}} e^{-\frac{z^2}{2}} \sinh \left( z\sqrt{\frac{3}{2}\beta^2D_e^2} + \frac{3}{2}\beta^2D_e^2 \right) \cosh^{R-2} \left( z\sqrt{\frac{3}{2}\beta^2D_e^2} + \frac{3}{2}\beta^2D_e^2 \right)}{\int \frac{dz}{\sqrt{2\pi}} e^{-\frac{z^2}{2}}\cosh^{R-1} \left( z\sqrt{\frac{3}{2}\beta^2D_e^2} + \frac{3}{2}\beta^2D_e^2 \right)}.
\end{eqnarray}
Through the transformation of the integration variable from $z$ to $z^\prime = z+\sqrt{\frac{3}{2}\beta^2D_e^2}$, this self-consistent equation turns out to be equivalent to that of overlaps, (\ref{eq:speq_ov_mf1}) with (\ref{eq:speq_ov_mf2}).
We also obtain 
\begin{eqnarray}
 -\beta g_{n,R} &=& -\frac{3}{4}\beta^2n(R-1)D_e^2 + n(R-1)\log 2 + \frac{1}{4}\beta^2 n + \frac{1}{4}\beta^2 n(R-1) \nonumber \\
 && - \beta^2 n(R-1) D_e^3 - \frac{1}{2}\beta^2n(R-1)(R-2)D_e^3 + n\log 2 \nonumber \\
 && + n\log \int \frac{dz}{\sqrt{2\pi}} e^{-\frac{z^2}{2}}\cosh^{R-1} \left( z\sqrt{\frac{3}{2}\beta^2D_e^2} + \frac{3}{2}\beta^2D_e^2 \right),
\end{eqnarray}
which is equal to $-\beta g_{n,R}$ in the calculation using overlaps, (\ref{eq:gnR_ov_mf}).
Thus, we conclude that these two methods calculating the Franz-Parisi potential give the same condition as the phase transition condition calculated with the 1RSB ansatz.

\section{} \label{cal}
In this section we calculate several quantities with the two-body cavity field solution (\ref{eq:2sol_p}) and (\ref{eq:2sol_ph}).
First, we calculate (\ref{eq:qA}) by using $Dz \equiv \frac{dz}{\sqrt{2\pi}} e^{-\frac{1}{2} z^2}$ and $\Theta_l(z_1, \cdots, z_l; h_*, h_{\mathrm{s}*}) \equiv \tanh^{-1} \left[ \tanh\left( \beta \left| J \right| \right) \prod_{j=1}^l \tanh \left( z_j\sqrt{\beta h_*} + \beta h_{\mathrm{s}*} \right) \right]$.
The result is
\begin{eqnarray}
 \fl A_u^{(r_1, \cdots, r_k)} &=& \prod_{\tau: \sigma^{(r_1)}\cdots \sigma^{(r_k)}=u} \sum_{\tau_{(1)}, \tau_{(2)}}e^{\sum_{j=1}^2 \left[ \sum_{r^\prime_1< r^\prime_2}^{r^\prime_1, r^\prime_2 \neq 0} \beta h_* \sigma_{(j)}^{(r^\prime_1)} \sigma_{(j)}^{(r^\prime_2)} + \sum_{r=1}^{R-1}\beta h_{\mathrm{s}*}\sigma_{(j)}^{(0)}\sigma_{(j)}^{(r)} \right]} \nonumber \\
 \fl && \qquad \times e^{J\sum_{r=0}^{R-1}\beta_0^{(r)}\sigma_{(1)}^{(r)}\sigma_{(2)}^{(r)}\sigma^{(r)}} \nonumber \\
 \fl &=& \prod_{\tau: \sigma^{(r_1)}\cdots \sigma^{(r_k)}=u} e^{-\beta h_* (R-1)} \sum_{\tau_{(1)}, \tau_{(2)}} e^{\sum_{j=1}^2 \sum_{r=1}^{R-1}\beta h_{\mathrm{s}*}\sigma_{(j)}^{(0)}\sigma_{(j)}^{(r)}} \nonumber \\
 \fl && \qquad \times \prod_{r=0}^{R-1}\left\{ \cosh\left( \beta_0^{(r)}J \right) + \sigma_{(1)}^{(r)}\sigma_{(2)}^{(r)}\sigma^{(r)} \sinh\left( \beta_0^{(r)}J \right) \right\} \nonumber \\
 \fl && \qquad \times \int \prod_{j=1}^2 Dz_j e^{\sum_{j=1}^2 z_j \sqrt{\beta h_*} \sum_{r=1}^{R-1}\sigma_{(j)}^{(r)}} \nonumber \\
 \fl &=& \prod_{\tau: \sigma^{(r_1)}\cdots \sigma^{(r_k)}=u} e^{-\beta h_* (R-1)} \sum_{\tau_{(1)}, \tau_{(2)}} \int \prod_{j=1}^2 Dz_j \nonumber \\
 \fl && \qquad \times \left\{ \cosh\left( \beta_0^{(0)}J \right) + \sigma_{(1)}^{(0)}\sigma_{(2)}^{(0)}\sigma^{(0)} \sinh\left( \beta_0^{(0)}J \right) \right\} \nonumber \\
 \fl && \qquad \times \prod_{r=1}^{R-1} \left[ \left\{ \cosh\left( \beta_0^{(r)}J \right) + \sigma_{(1)}^{(r)}\sigma_{(2)}^{(r)}\sigma^{(r)} \sinh\left( \beta_0^{(r)}J \right) \right\} \right. \nonumber \\
 \fl && \qquad \times \left. e^{\sum_{j=1}^2 \beta h_{\mathrm{s}*}\sigma_{(j)}^{(0)}\sigma_{(j)}^{(r)}} e^{\sum_{j=1}^2 z_j \sqrt{\beta h_*} \sigma_{(j)}^{(r)}} \right] \nonumber \\
 \fl &=& \prod_{\tau: \sigma^{(r_1)}\cdots \sigma^{(r_k)}=u} e^{-\beta h_* (R-1)} \nonumber \\
 \fl && \qquad \times \int \prod_{j=1}^2 Dz_j \sum_{\sigma_{(1)}^{(0)}, \sigma_{(2)}^{(0)}} \left\{ \cosh\left( \beta^\prime J \right) + \sigma_{(1)}^{(0)}\sigma_{(2)}^{(0)}\sigma^{(0)} \sinh\left( \beta^\prime J \right) \right\} \nonumber \\
 \fl && \qquad \times \prod_{r=1}^{R-1} \left[ \cosh(\beta J) \prod_{j=1}^2 2\cosh\left( z_j\sqrt{\beta h_*} + \beta h_{\mathrm{s}*}\sigma_{(j)}^{(0)} \right) \right. \nonumber \\
 \fl && \qquad \qquad \left. + \sigma^{(r)}\sinh(\beta J) \prod_{j=1}^2 2\sinh\left( z_j\sqrt{\beta h_*} + \beta h_{\mathrm{s}*}\sigma_{(j)}^{(0)} \right) \right] \nonumber \\
 \fl &=& \prod_{\tau: \sigma^{(r_1)}\cdots \sigma^{(r_k)}=u} 2^R e^{-\beta h_* (R-1)} \left\{ \prod_{r=0}^{R-1} \cosh\left( \beta_0^{(r)} J \right) \right\} \nonumber \\
 \fl && \qquad \times \int \prod_{j=1}^2 Dz_j \left\{ \prod_{j=1}^2 \cosh^{R-1} \left( z_j\sqrt{\beta h_*} + \beta h_{\mathrm{s}*} \right) \right\} \nonumber \\
 \fl && \qquad \times \left[ \left\{ 1+\sigma^{(0)}\tanh\left( \beta^\prime J \right) \right\} \prod_{r=1}^{R-1} \left\{ 1+ \sigma^{(r)}\mathrm{sgn}(J)\tanh\Theta_2 \right\} \right. \nonumber \\
 \fl && \qquad \qquad \left. + \left\{ 1-\sigma^{(0)}\tanh\left( \beta^\prime J \right) \right\} \prod_{r=1}^{R-1} \left\{ 1- \sigma^{(r)}\mathrm{sgn}(J)\tanh\Theta_2 \right\} \right] \nonumber \\
 \fl &=& \prod_{\tau: \sigma^{(r_1)}\cdots \sigma^{(r_k)}=u} 2^R e^{-\beta h_* (R-1)} \left\{ \prod_{r=0}^{R-1} \cosh\left( \beta_0^{(r)} J \right) \right\} \nonumber \\
 \fl && \qquad \times \int \prod_{j=1}^2 Dz_j \left\{ \prod_{j=1}^2 \cosh^{R-1} \left( z_j\sqrt{\beta h_*} + \beta h_{\mathrm{s}*} \right) \right\} \nonumber \\
 \fl && \qquad \times \sum_{\nu=\pm 1} \left\{ 1+ \nu\sigma^{(0)}\tanh\left( \beta^\prime J \right) \right\} \prod_{r=1}^{R-1} \left\{ 1+ \nu\sigma^{(r)}\mathrm{sgn}(J)\tanh\Theta_2 \right\}.
\end{eqnarray}

We also calculate $w\left( \mathfrak{h} \right)$ in (\ref{eq:def_w}) as
\begin{eqnarray}
 \fl w\left( \mathfrak{h} \right) &=& \sum_\tau e^{\sum_{r=1}^{R-1}\beta h_{\mathrm{s}*}\sigma^{(0)}\sigma^{(r)} + \sum_{r_1< r_2}^{r_1, r_2 \neq 0} \beta h_* \sigma^{(r_1)}\sigma^{(r_2)}} \nonumber \\
 \fl &=& \sum_\tau e^{\sum_{r=1}^{R-1}\beta h_{\mathrm{s}*}\sigma^{(0)}\sigma^{(r)} - \frac{1}{2}\beta h_* (R-1)} \int Dz e^{z\sqrt{\beta h_*}\sum_{r=1}^{R-1}\sigma^{(r)}} \nonumber \\
 \fl &=& \sum_{\sigma^{(0)}} e^{- \frac{1}{2}\beta h_* (R-1)} \int Dz 2^{R-1} \cosh^{R-1} \left( z\sqrt{\beta h_*} + \beta h_{\mathrm{s}*}\sigma^{(0)} \right) \nonumber \\
 \fl &=& 2^{R-1} e^{- \frac{1}{2}\beta h_* (R-1)} \int Dz \left\{ \cosh^{R-1} \left( z\sqrt{\beta h_*} + \beta h_{\mathrm{s}*} \right) + \cosh^{R-1} \left( z\sqrt{\beta h_*} - \beta h_{\mathrm{s}*} \right) \right\} \nonumber \\
 \fl &=& 2^R e^{- \frac{1}{2}\beta h_* (R-1)} \int Dz \cosh^{R-1} \left( z\sqrt{\beta h_*} + \beta h_{\mathrm{s}*} \right).
\end{eqnarray}

\section{} \label{AC}
In this section, we derive the identities (\ref{eq:identity1}) and (\ref{eq:identity2}).
First, by defining
\begin{eqnarray}
 \phi(n) &\equiv& \sum_{k=0}^n \binom{n}{k} f(k),
\end{eqnarray}
we obtain
\begin{eqnarray}
 \phi(n+1) &=& f(n+1) + f(0) + \sum_{k=1}^n \binom{n+1}{k} f(k) \nonumber \\
 &=& f(n+1) + f(0) + \sum_{k=1}^n \left( \binom{n}{k} + \binom{n}{k-1} \right) f(k) \nonumber \\
 &=& f(n+1) + \sum_{k=0}^n \binom{n}{k} f(k) + \sum_{k=0}^{n-1} \binom{n}{k} f(k+1) \nonumber \\
 &=& \phi(n) + \sum_{k=0}^{n} \binom{n}{k} f(k+1).
\end{eqnarray}
Similarly, by introducing
\begin{eqnarray}
 \tilde{\phi}^{(m)}(n) &\equiv& \sum_{k=0}^n \binom{n}{k} f(k+m), 
\end{eqnarray}
we also obtain
\begin{eqnarray}
 \tilde{\phi}^{(0)}(n) &=& \phi(n) \\
 \tilde{\phi}^{(0)}(n+1) &=& \tilde{\phi}^{(0)}(n) + \tilde{\phi}^{(1)}(n) \\
 \tilde{\phi}^{(m)}(n+1) &=& \tilde{\phi}^{(m)}(n) + \tilde{\phi}^{(m+1)}(n).
 \label{eq:phi_til_m_eq}
\end{eqnarray}
By considering the sum $\sum_{m=0}^\infty (-1)^m \tilde{\phi}^{(m)}(n+1)$, we obtain
\begin{eqnarray}
 && \sum_{m=0}^\infty (-1)^m \tilde{\phi}^{(m)}(n+1) \nonumber \\
 &=& \sum_{m=0}^\infty (-1)^m \tilde{\phi}^{(m)}(n) + \sum_{m=0}^\infty (-1)^m \tilde{\phi}^{(m+1)}(n) \nonumber \\
 &=& \tilde{\phi}^{(0)}(n) + \sum_{m=1}^\infty (-1)^m \tilde{\phi}^{(m)}(n) + \sum_{m=0}^\infty (-1)^m \tilde{\phi}^{(m+1)}(n) \nonumber \\
 &=& \phi(n) + (-1)\sum_{m=0}^\infty (-1)^m \tilde{\phi}^{(m+1)}(n) + \sum_{m=0}^\infty (-1)^m \tilde{\phi}^{(m+1)}(n) \nonumber \\
 &=& \phi(n).
\end{eqnarray}
Consequently, we obtain the relation
\begin{eqnarray}
 \phi(n) &=& \sum_{m=0}^\infty (-1)^m \tilde{\phi}^{(m)}(n+1).
\end{eqnarray}
Particularly, in the limit $n\rightarrow -1$, we obtain the identity (\ref{eq:identity1}):
\begin{eqnarray}
 \lim_{n\rightarrow -1} \phi(n) &=& \sum_{m=0}^\infty (-1)^m \tilde{\phi}^{(m)}(0) \nonumber \\
 &=& \sum_{m=0}^\infty (-1)^m f(m) \nonumber \\
 &=& \left. \frac{\partial }{\partial x} \sum_{m=0}^\infty (-1)^m x^{f(m)} \right|_{x=1}.
\end{eqnarray}
Similarly, by considering the sum $\sum_{m=0}^\infty (-1)^m (m+1) \tilde{\phi}^{(m)}(n+2)$ and using the identity $\tilde{\phi}^{(m)}(n+2) = \tilde{\phi}^{(m)}(n) + 2\tilde{\phi}^{(m+1)}(n) + \tilde{\phi}^{(m+2)}(n)$, we also obtain the identity (\ref{eq:identity2}):
\begin{eqnarray}
 \lim_{n\longrightarrow -2} \sum_{k=0}^n \binom{n}{k} f(k) &=& \left. \frac{\partial }{\partial x} \sum_{m=0}^\infty (-1)^m (m+1) x^{f(m)} \right|_{x=1}.
\end{eqnarray}

We remark that when $f(k)$ depends on $n$, we just have to consider $\phi_{n_1}(n_2) \equiv \sum_{k=0}^{n_2} \binom{n_2}{k} f(k;n_1)$.
By applying the same argument to $n_2$ and taking $n_1=n_2$, we finally obtain the identity
\begin{eqnarray}
 \lim_{n_2\rightarrow -1} \phi_{n_2}(n_2) = \left. \frac{\partial }{\partial x} \sum_{m=0}^\infty (-1)^m x^{f(m;-1)} \right|_{x=1}.
\end{eqnarray}
It should also be noted that our identities (\ref{eq:identity1}) and (\ref{eq:identity2}) can also be derived in terms of the gamma function \cite{DotMez1997}.

We can apply these formulas to our $\Phi_{a,b}$.
By defining $\Theta_{l(0)}(z_1, \cdots, z_l) \equiv \Theta_l(z_1, \cdots, z_l; h_{*(0)}, h_{\mathrm{s}*(0)})$ and
\begin{eqnarray}
 \fl B_{(0)}(m) &\equiv& \log \int \prod_{j=1}^2 Dz_j \sum_{\nu=\pm 1} \left\{ 1+ \nu\tanh\left( \beta^\prime \left| J \right| \right) \right\} \left\{ \frac{1- \nu\tanh\Theta_{2(0)}}{1+ \nu\tanh\Theta_{2(0)}} \right\}^m,
 \label{eq:B0m}
\end{eqnarray}
we obtain in the limit $R \rightarrow 1$
\begin{eqnarray}
 \Phi^{(0)}_{1,1} &\equiv& \left. \frac{\partial }{\partial x} \sum_{m=0}^\infty (-1)^m x^{B_{(0)}(m)} \right|_{x=1},
 \label{eq:Phi11} \\
 \Phi^{(0)}_{1,2} &\equiv& \left. \frac{\partial }{\partial x} \sum_{m=0}^\infty (-1)^m x^{B_{(0)}(-m)} \right|_{x=1},
 \label{eq:Phi12} \\
 \Phi^{(0)}_{2,1} &\equiv& \left. \frac{\partial }{\partial x} \sum_{m=0}^\infty (-1)^m (m+1) x^{B_{(0)}(m)} \right|_{x=1},
 \label{eq:Phi21} \\
 \Phi^{(0)}_{2,2} &\equiv& \left. \frac{\partial }{\partial x} \sum_{m=0}^\infty (-1)^m (m+1) x^{B_{(0)}(-m)} \right|_{x=1},
 \label{eq:Phi22} \\
 \Phi^{(0)}_{2,3} &\equiv& \left. \frac{\partial }{\partial x} \sum_{m=0}^\infty (-1)^m (m+1) x^{B_{(0)}(m+1)} \right|_{x=1},
 \label{eq:Phi23} \\
 \Phi^{(0)}_{2,4} &\equiv& \left. \frac{\partial }{\partial x} \sum_{m=0}^\infty (-1)^m (m+1) x^{B_{(0)}(-m-1)} \right|_{x=1}.
 \label{eq:Phi24}
\end{eqnarray}


\end{document}